\documentclass[twocolumn,preprintnumbers,amsmath,amssymb,superscriptaddress]{revtex4}

\usepackage{times}
\usepackage[pdftex]{graphicx}
\usepackage[pdftex]{epsfig}
\usepackage{amsmath}
\usepackage{pslatex}
\usepackage{amssymb}
\usepackage{amsfonts}
\usepackage{color}
\usepackage{wrapfig}
\usepackage{eucal}
\usepackage{hhline}
\usepackage{threeparttable}
\usepackage{supertabular}
\usepackage{multirow}
\usepackage{tabularx}

\newcommand{\opd}[2]{\mbox{$\hat{#1}_{#2}^{\dagger}$}}
\newcommand{\op}[2]{\mbox{$\hat{#1}_{#2}$}}

\def\be{\begin{equation}}
\def\ee{\end{equation}}
\def\bea{\begin{eqnarray}}
\def\eea{\end{eqnarray}}

\newcommand{\avg}[1]{\mbox{$\langle#1\rangle$}}

\newcommand{\opdagger}[2]{\mbox{$\hat{#1}_{#2}^{\dagger}$}}

\newcolumntype{Y}{>{\centering\arraybackslash}X}

\begin{document}

\pagenumbering{arabic}

\title{Coherent optical wavelength conversion via cavity-optomechanics}

\author{Jeff T. Hill}
\thanks{These authors contributed equally to this work}
\author{Amir H. Safavi-Naeini}
\thanks{These authors contributed equally to this work}
\author{Jasper Chan}
\affiliation{Thomas J. Watson, Sr., Laboratory of Applied Physics, California Institute of Technology, Pasadena, CA 91125}
\author{Oskar Painter}
\email{opainter@caltech.edu}
\homepage{http://copilot.caltech.edu}
\affiliation{Thomas J. Watson, Sr., Laboratory of Applied Physics, California Institute of Technology, Pasadena, CA 91125}

\date{\today}

\begin{abstract}
We theoretically propose and experimentally demonstrate coherent wavelength conversion of optical photons using photon-phonon translation in a cavity-optomechanical system.  For an engineered silicon optomechanical crystal nanocavity supporting a $4$~GHz localized phonon mode, optical signals in a $1.5$~MHz bandwidth are coherently converted over a $11.2$~THz frequency span between one cavity mode at wavelength $1460$~nm and a second cavity mode at $1545$~nm with a 93\% internal (2\% external) peak efficiency.  The thermal and quantum limiting noise involved in the conversion process is also analyzed, and in terms of an equivalent photon number signal level are found to correspond to an internal noise level of only 6 and $4 \times 10^{-3}$ quanta, respectively.
\end{abstract}

\maketitle


The interaction of light with acoustic and molecular mechanical vibrations enables a great many optical functions used in communication systems today, such as amplification, modulation, wavelength conversion, and switching~\cite{Toulouse2005}.  Conventionally, these functions are realized in long (centimeter to many meter) waveguide based devices, relying on inherent materials' properties and requiring intense optical pump beams.  With the technological advancements in the fields of nanomechanics and nanophotonics, it is now possible to engineer interactions of light and mechanics.  Progress in this area has included enhanced nonlinear optical interactions in structured silica fibers~\cite{Kang2009}, near quantum-limited detection of nanomechanical motion~\cite{Teufel2011}, and the radiation pressure cooling of a mesoscopic mechanical resonator to its quantum ground state of motion~\cite{Teufel2011,Chan2011}.  Coupling of electromagnetic and mechanical degrees of freedom, in which the coherent interaction rate is larger than the thermal decoherence rate of the system, as realized in the ground-state cooling experiments, opens up an array of new applications in classical and quantum optics.  This realization, along with the inherently broadband nature of radiation pressure, has spawned a variety of proposals for converting between photons of disparate frequencies~\cite{Tian2010,Stannigel2010,Safavi-Naeini2011b,Wang2011} through interaction with a mechanical degree of freedom - proposals which are but one of many possible expressions of hybrid optomechanical systems~\cite{Wallquist2009}. 

The ability to coherently convert photons between disparate wavelengths has broad technological implications not only for classical communication systems, but also future quantum networks~\cite{Kimble2008,Kielpinski2011,Ritter2012}.  For example, hybrid quantum networks require a low loss interface capable of maintaining quantum coherence while connecting spatially separate systems operating at incompatible frequencies~\cite{Wallquist2009}.  For this reason, photons operating in the low loss telecommunications band are often proposed as a conduit for connecting different physical quantum systems~\cite{Riedmatten2008}.  It has also been realized that a wide variety of quantum systems lend themselves to coupling with mechanical elements.  A coherent interface between mechanics and optics, then, could provide the required quantum links of a hybrid quantum network~\cite{Stannigel2010}.  Until now, most experiments demonstrating either classical or quantum wavelength conversion have used intrinsic optical nonlinearities of materials~\cite{Toulouse2005,Huang1992,Tanzilli2005}. In this Letter we show that by patterning an optomechanical crystal (OMC) nanobeam from a thin film of silicon, engineered GHz mechanical resonances can be used to convert photons within a $1.5$~MHz bandwidth coherently over an optical frequency span of 11.2~THz, at internal efficiencies exceeding 90\%, and with a thermal(quantum)-limited noise of only 6 ($4 \times 10^{-3}$) quanta.  Such cavity optomechanical systems are not just limited to the optical frequency domain, but may also find application to the interconversion of microwave and optical photons\cite{Safavi-Naeini2011b,Regal2011}, enabling a quantum-optical interface to superconducting quantum circuits\cite{Schoelkopf2008}. 

\begin{figure*}[ht!]
\begin{center}
\includegraphics[width=1.9\columnwidth]{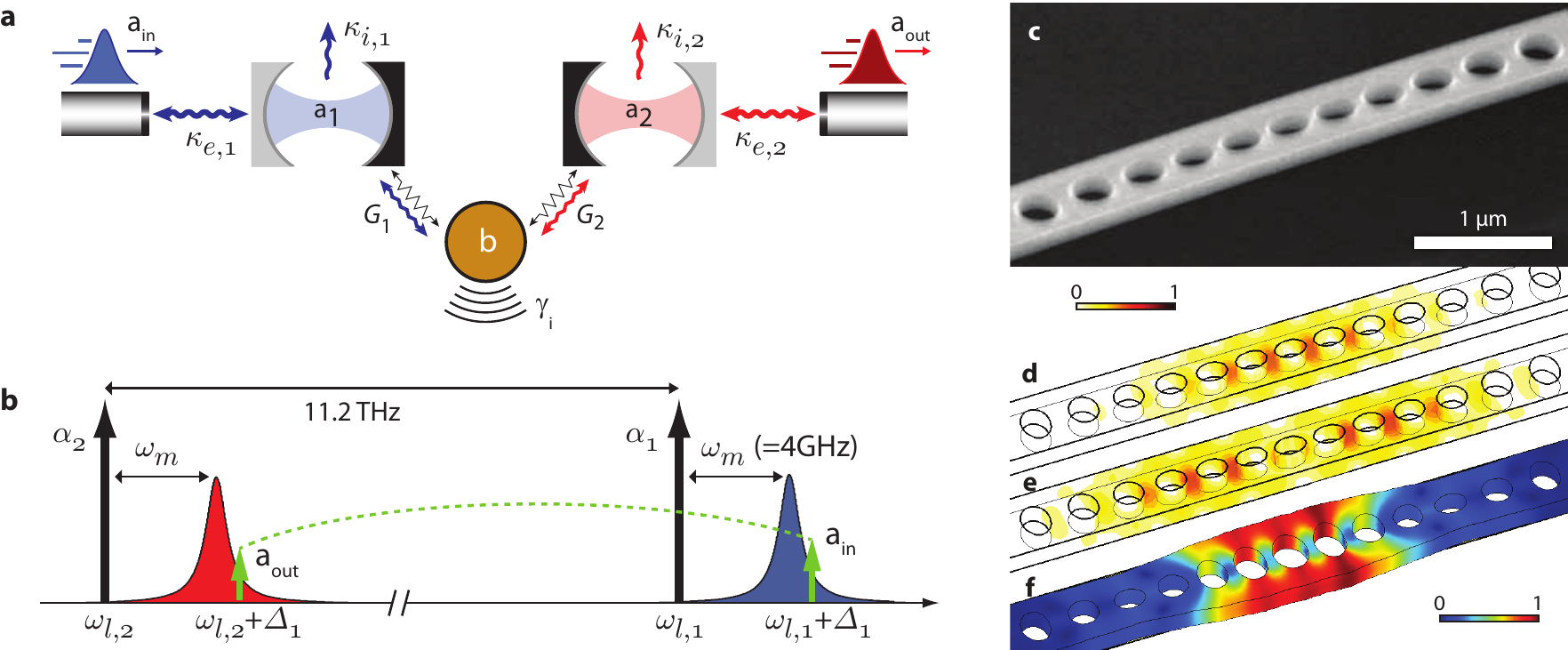}
\caption{\textbf{System model and physical realization.} \textbf{a}, Diagram of the wavelength conversion process as realized via two separate Fabry-Perot cavities.  The two optical cavity modes, $a_1$ and $a_2$, are coupled to the same mechanical mode, $b$, with coupling strengths $G_1$ and $G_2$, respectively.  The optical cavity modes are each coupled to an external waveguide (with coupling strengths $\kappa_{e,1}, \kappa_{e,2}$), through which optical input and output signals are sent.  The optical cavities also have parasitic (intrinsic) loss channels, labeled $\kappa_{i,1}$ and $\kappa_{i,2}$, whereas the mechanical mode is coupled to its thermal bath at rate $\gamma_i$. \textbf{b}, Schematic indicating the relevant optical frequencies involved in the wavelength conversion process.  The cavity control laser beams, labeled $\alpha_1$ and $\alpha_2$, are tuned a mechanical frequency red of the corresponding optical cavity resonances.  An input signal ($a_{\text{in}}$) is sent into the input cavity at frequency $\omega_{l,1} + \Delta_1$.  The input signal is converted into an output signal ($a_{\text{out}}$) at frequency $\omega_{l,2} + \Delta_1$ via the optomechanical interaction. \textbf{c}, Scanning electron micrograph (SEM) of the fabricated silicon nanobeam optomechanical cavity.  \textbf{d}, Finite element method (FEM) simulation of the electromagnetic energy density of the first and, \textbf{e}, second order optical cavity modes of the silicon nanobeam.  \textbf{f}, FEM simulation of the displacement field of the co-localized mechanical mode.}
\label{fig:concept}
\end{center}
\end{figure*}


Conceptually, the proposed system for wavelength conversion can be understood from Fig.~\ref{fig:concept}a.  Two optical cavity modes $\op{a}{k}$ ($k=1,2$) are coupled to a common mechanical mode $\op{b}{}$ via an interaction Hamiltonian $H = \sum_{k} \hbar g_{k} \opd{a}{k}\op{a}{k} (\op{b}{} + \opd{b}{})$, where $\op{a}{k}$, $\op{b}{}$ are the annihilation operators and $g_k$ is the optomechanical coupling rate between the mechanical mode and the $k$th cavity mode.  Physically, $g_{k}$ represents the frequency shift of cavity mode $k$ due to the zero-point motion of the mechanical resonator.  Wavelength conversion is driven by two control laser beams ($\alpha_k$ in Fig.~\ref{fig:concept}b), of frequency $\omega_{l,k}$ and nominal detuning $\delta_k \equiv \omega_{k} - \omega_{l,k} = \omega_{m}$ to the red of cavity resonance at frequency $\omega_{k}$.   In the resolved sideband regime, where $\omega_m \gg \kappa_k$ ($\kappa_k$ the bandwidth of the $k$th cavity mode), the spectral filtering of each cavity preferentially enhances photon-phonon exchange.  The resulting beam-splitter-like Hamiltonian is $H = \sum_{k} \hbar G_{k}(\opd{a}{k} \op{b}{} + \op{a}{k} \opd{b}{})$~\cite{Aspelmeyer2010,Safavi-Naeini2011b}, where $G_{k} \equiv g_k\sqrt{n_{\mathrm{c},k}}$ is the parametrically enhanced optomechanical coupling rate due to the $\alpha_{k}$ control beam ($n_{\mathrm{c},k}$ is the control-beam induced intracavity photon number).  In the weak-coupling limit, $G_k \ll \kappa_k$, this interaction effectively leads to an additional mechanical damping rate, $\gamma_{\mathrm{OM},k} = 4G_k^{2}/{\kappa_{k}}$.  The degree to which this optomechanical loss rate dominates the intrinsic mechanical loss is called the cooperativity, $C_{k} = \gamma_{\mathrm{OM},k}/\gamma_i$.  At large cooperativities, the optomechanical damping has been used as a nearly noiseless loss channel to cool the mechanical mode to its ground state~\cite{Teufel2011,Chan2011}.  In the case of a single optical cavity system, it has also been used as a coherent channel allowing inter-conversion of photons and phonons leading to the observation of electromagnetically-induced transparency (EIT)~\cite{Weis2010,Safavi-Naeini2011}.  In the double optical cavity system presented here, coherent wavelength conversion of photons results.


As shown in Fig.~\ref{fig:concept}a, each optical cavity is coupled not just to a common mechanical mode, but also to an optical bath at rate $\kappa_{i,k}$ and to an external photonic waveguide at rate $\kappa_{e,k}$ (the total cavity linewidth is $\kappa_k=\kappa_{i,k}+\kappa_{e,k}$).  The external waveguide coupling provides an optical interface to the wavelength converter, and in this work consists of a single-transverse-mode waveguide, bi-directionally coupled to each cavity mode.  The efficiency of the input/output coupling is defined as $\eta_k = \kappa_{\mathrm{e},k}/2\kappa_k$, half that of the total bi-directional rate.  Although the wavelength converter operates symmetrically, here we will designate the higher frequency cavity mode ($k=1$) as the input cavity and the lower frequency cavity ($k=2$) as the output cavity.  As shown in Fig.~\ref{fig:concept}b, photons sent into the wavelength converter with detuning $\Delta_{1} \sim \omega_{m}$ from the control laser $\alpha_1$, are converted to photons $\Delta_{1}$ detuned from the control laser $\alpha_2$, an $11.2$~THz frequency span for the device studied here. 

\begin{figure*}[ht!]
\begin{center}
\includegraphics[width=2\columnwidth]{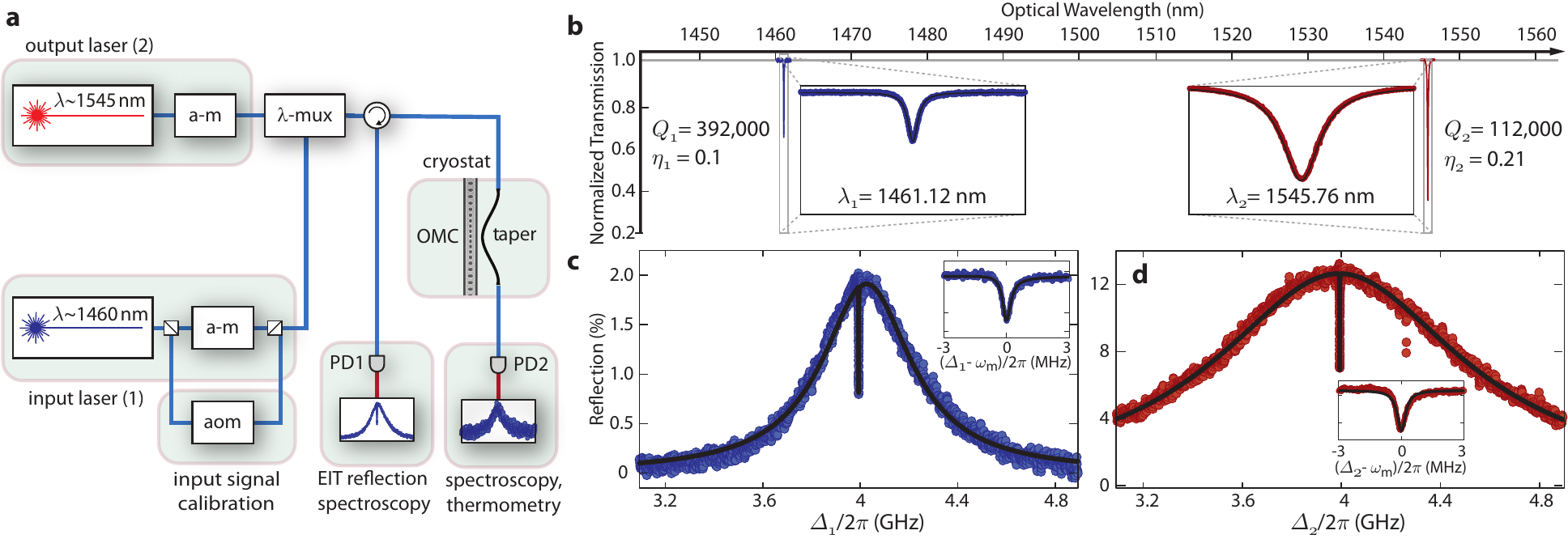}
\caption{\textbf{Experimental set-up and optical spectroscopy.} \textbf{a}, Two tunable external cavity diode lasers are used as control beams driving the wavelength conversion process. The input (output) control laser is locked a mechanical frequency red-detuned from the first-order (second-order) cavity mode at $\lambda\approx1460$~nm ($\lambda\approx1545$~nm).  Both control beams can be amplitude modulated (a-m) to perform EIT-like spectroscopy of the cavity modes, or in the case of the input laser, to generate the input sideband signal for the wavelength conversion process.  As described in the Appendix, an acousto-optic modulator (aom) is used to calibrate the input sideband signal.  The light from both lasers is combined using a wavelength multiplexer ($\lambda$-mux), and then sent into a dimpled optical fiber taper which is coupled to the OMC cavity.  The cavity sample is placed in a cryostat to pre-cool it down to $T\approx14$~K.  The transmitted light from the cavity is sent to a high-speed photodetector (PD2) which is connected to a spectrum analyzer to measure the converted signal on the output laser.  The reflected optical signal from the cavity is directed via an optical circulator to a second high-speed photodetector (PD1) to probe the EIT-like spectrum of each cavity mode.  \textbf{b}, Broad wavelength transmission scan showing fiber taper coupling to both the first-order ($\lambda\approx1460$~nm) and second-order ($\lambda\approx1545$~nm) cavity modes. \textbf{c}, EIT scan of cavity mode $\op{a}{}_1$, with inset showing a zoomed-in scan of the transparency window.  The solid black lines correspond to fits to a theoretical model allowing extraction of system parameters ($\kappa_{1}$, $\gamma$, $\delta_{1}$, and $\omega_m$).  \textbf{d}  Corresponding EIT scan of cavity mode $\op{a}{}_2$.   For measurements in both \textbf{c} and \textbf{d} the control beam intensities were set such that $\gamma_{\text{OM},1}\approx\gamma_{\text{OM},2}\gg\gamma_i$, with detunings $\delta_1\gtrsim\omega_m$ and $\delta_2\approx\omega_m$.}
\label{fig:EIT}
\end{center}
\end{figure*}


The details of the conversion process can be understood by solving the Heisenberg-Langevin equations (see Appendix).  We linearize the system and work in the frequency domain, obtaining through some algebra the scattering matrix element $s_{21}(\omega)$, which is the complex, frequency-dependent conversion coefficient between the input field at cavity $\op{a}{}_1$, and the output field at cavity $\op{a}{}_2$. This coefficient is given by the expression
\be
\label{eqn:skj}
s_{21}(\omega) = \sqrt{\eta_2 \eta_1}  \frac{\sqrt{\gamma_{\mathrm{OM},2}\gamma_{\mathrm{OM},1}}}{i(\omega_m - \omega) + \gamma/2},
\ee
where $\gamma = \gamma_i + \gamma_{\mathrm{OM},2} + \gamma_{\mathrm{OM},1}$ is the total mechanical damping rate and equal to the bandwidth of the conversion process.
From this expression, the spectral density of a converted signal $S_{\mathrm{out},2}(\omega)$, given the input signal spectral density $S_{\mathrm{in},1}(\omega)$, may be found and is given by
\bea
S_{\mathrm{out},2}(\omega) &=&  \eta_2 \eta_1 \frac{\gamma_{\text{OM},2} \gamma_{\text{OM},1}}{(\omega + \omega_m)^2 + (\gamma/2)^2}(n_\text{added} + S_{\mathrm{in},1}(\omega)). \label{eqn:psd_expr}
\eea
These spectral densities have units of photons/Hz$\cdot$s and are proportional to optical power. The added noise, $n_\text{added}$, arises from thermal fluctuations of the mechanical system and the quantum back-action noise of light present in each optical mode.
From here, we see that in a system with ideal cavity-waveguide coupling, ($\eta_1,\eta_2=1$), the peak internal photon conversion efficiency is given by
\be
\eta_\text{max,int} = \frac{4C_1 C_2}{(1+C_1+C_2)^2}.
\ee
This efficiency only depends on the internal coupling of the optomechanical system, and for both $C_1 = C_2$ and $C_1, C_2 \gg 1$, approaches unity.  The latter condition can be understood from requiring the coupling between the optical and mechanical modes to overtake the intrinsic mechanical loss rate, while the first requirement is due to impedance-like matching~\cite{Safavi-Naeini2011b}. The total system efficiency is $\eta_\text{max} = \eta_1 \eta_2 \eta_\text{max,int}$.


The optomechanical system used in this work, shown in Fig.~\ref{fig:concept}c-f, consists not of a separate set of optical cavities, but rather of a single optomechanical crystal (OMC) nanobeam cavity.  The OMC nanobeam is fabricated from a silicon-on-insulator (SOI) microchip~\cite{Eichenfield2009a}, in which the top, $220$~nm thick, silicon device layer is patterned with a quasi-periodic linear array of etched air holes.  The larger air holes on either end of the beam induce Bragg-like reflection of both guided optical and acoustic waves, resulting in strongly confined optical and mechanical resonances at the beam's center.  The device used in this work is designed to have both a first-order ($\op{a}{1},\, \lambda\approx1460~\mathrm{nm}$) and a second-order ($\op{a}{2},\, \lambda\approx1545~\mathrm{nm}$) optical cavity resonance of high quality factor.  Both of these optical cavity modes are dispersively coupled to the same GHz-frequency mechanical resonance, depicted in Fig.~\ref{fig:concept}f ($\op{b}{},\,\omega_{m}/2\pi = 3.993~\mathrm{GHz}, Q_m = 87 \times 10^3$ at $T \approx 14~\mathrm{K}$). A tapered optical fiber waveguide, placed in close proximity ($\sim 100$~nm) to the OMC nanobeam cavity~\cite{Michael2007} using precision stages, is used to evanescently couple light into and out of the cavity modes.

A schematic of the experimental set-up used to characterize the OMC wavelength converter is shown in Fig.~\ref{fig:EIT}a.  Measurements were performed in vacuum and at low temperature inside a continuous flow cryostat with a cold finger temperature of $9$~K (the corresponding sample temperature, as inferred from the thermal bath temperature of the localized mechanical mode of the OMC cavity, is $14$~K~\cite{Chan2011}).  Initial characterization of the optical cavity modes is performed by scanning the tunable control lasers across a wide bandwidth and recording the transmitted optical intensity through the optical fiber taper waveguide.  Such a wavelength scan is shown in Fig.~\ref{fig:EIT}b, from which the resonance frequency ($\omega_{1}/2\pi = 205.3~\mathrm{THz}$, $\omega_{2}/2\pi = 194.1~\mathrm{THz}$), cavity linewidth ($\kappa_1/2\pi=520$~MHz, $\kappa_2=1.73$~GHz), and waveguide coupling efficiency ($\eta_1=0.10$, $\eta_2=0.21$) of the two nanobeam cavity modes are determined.  

Further characterization of the optomechanical cavity is performed by using the control laser beams in conjunction with a weak sideband probe.  With control beams $\alpha_1$ and $\alpha_2$ detuned a mechanical frequency to the red of their respective cavity modes ($\delta_{1,2}=\omega_m$), a weak sideband signal generated from $\alpha_1$ is swept across the first-order cavity mode.  The resulting reflected sideband signal versus sideband frequency shift $\Delta_1$ is plotted in Fig.~\ref{fig:EIT}c, showing the broad cavity resonance along with a narrow central reflection dip (see Fig.~\ref{fig:EIT}c inset).  The narrow reflection dip, akin to the EIT transparency window in atomic systems, is due to the interference between light coupled directly into the cavity mode and light coupled indirectly through the mechanical mode~\cite{Weis2010,Safavi-Naeini2011}.  A similar EIT response is shown in Fig.~\ref{fig:EIT}d for the second-order cavity mode.  Each of the transparency windows occur at $\Delta_{k}=\omega_m$, with a bandwidth equal to the optically damped mechanical linewidth ($\gamma=\gamma_i+\gamma_{\text{OM},1}+\gamma_{\text{OM},2}$).  By fitting the optical resonance lineshape and transparency windows to theory, one can also extract the control beam detunings $\delta_k$.  In what follows we use this sort of EIT reflection spectroscopy, time-multiplexed in between wavelength conversion measurements, to set and stabilize the frequency of the control beams to $\delta_k=\omega_m$.



\begin{figure}[t]
\begin{center}
\includegraphics[width=\columnwidth]{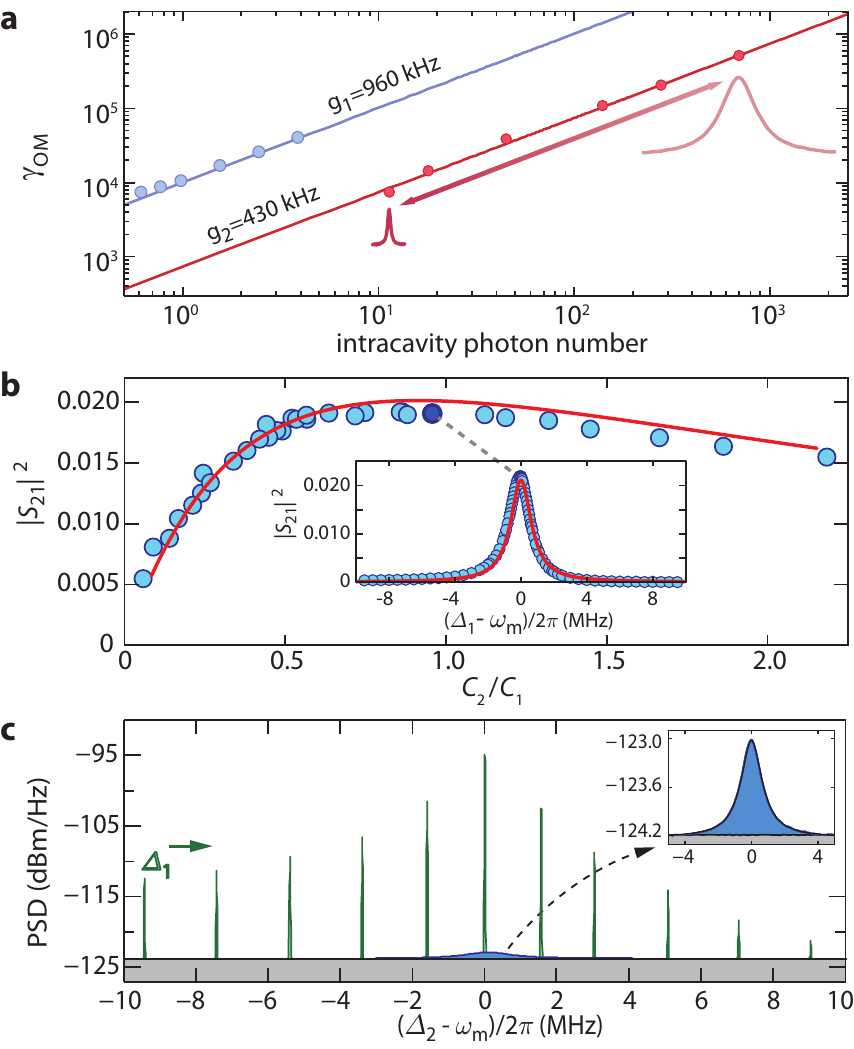}
\caption{\textbf{Wavelength conversion efficiency and bandwidth.} \textbf{a}, Plot of the optically induced mechanical damping versus control beam intensity (intracavity photon number) for each cavity mode, as measured through the mechanical thermal noise spectrum imprinted on the optical output intensity of $\op{a}{}_2$.  The calibration curve for cavity mode $\op{a}{}_2$ is performed with $\alpha_1$ turned off, whereas the curve for cavity mode $\op{a}{}_1$ is generated with a weak $\alpha_2$ such that $\gamma_{\text{OM},2}\ll\gamma_{\text{OM},1}$ for all measured points.   \textbf{b}, End-to-end power conversion efficiency of input signal to output signal for $\Delta_1=0$ as a function of the ratio of the cooperativities of the control beams.  In this plot the control beam intensity for the first-order cavity mode is held fixed (with $C_1 \approx 16$), while the intensity of the control beam of the second-order cavity mode is swept from $C_2 \ll C_1$ to $C_2 > C_1$.  The blue circles correspond to measured data points, whereas the solid red line is a theoretical curve using independently measured system parameters. Inset shows the conversion efficiency versus input signal detuning for matched control beams $C_1=C_2\approx16$, indicating a conversion bandwidth of $\sim 1.55~\textrm{MHz}$.  \textbf{c},  Measured power spectrum of the output optical channel for $C_1 = C_2 \approx 16$, showing a series ($\Delta_1$ sweep) of converted input tones (green) sitting on top of a much smaller thermal noise pedestal (blue).  Inset shows a zoom-in of the thermal noise pedestal.}
\label{fig:results}
\end{center} 
\end{figure}


Coherent wavelength conversion can be thought of as occurring between the input transparency window of $\op{a}{1}$ and the output transparency window of $\op{a}{2}$, with the phonon transition mediating the conversion.  As shown above, conversion efficiency is theoretically optimized for matched cooperativities of control beam $\alpha_1$ and $\alpha_2$ (or equivalently $\gamma_{\text{OM,1}}=\gamma_{\text{OM,2}}$).  Calibration of the optically induced mechanical damping by the control beams can be performed as in the EIT spectroscopy described above, or by measuring the spectral content of the photodetected transmission intensity of $\op{a}{2}$ in the absence of any optical input.  Such a measurement (see set-up in Fig.~\ref{fig:EIT}a and Appendix) measures the mechanical resonator linewidth through the noise spectrum generated by the beating of noise sideband photons generated by thermal motion of the mechanical resonator with the control beam $\alpha_2$.  Figure~\ref{fig:results}a plots the inferred optomechanically induced damping of the mechanical resonator as a function of the power (intracavity photon number, $n_c$) of control beams $\alpha_1$ and $\alpha_2$, the slope of which gives the zero-point optomechanical coupling for both cavity modes ($g_1/2\pi = 960~\mathrm{kHz}$ and $g_2/2\pi = 430~\mathrm{kHz}$).

To quantify the efficiency of the wavelength conversion process the input cavity control beam $\alpha_1$ is held fixed at a detuning $\delta_1=\omega_m$ and a power producing an intracavity photon population of $n_{\mathrm{c},1} = 100$, corresponding to large cooperativity $C_1 \approx 16$.  The input signal, $a_{in}$, is generated as an upper sideband of $\alpha_1$ using an electro-optic intensity modulator (the lower sideband is rejected by the wavelength converter as it is detuned to $\Delta_1\approx-\omega_m$).  Conversion of the $a_{in}$ sideband to an output signal emanating from cavity mode $\op{a}{}_2$ is completed by applying control beam $\alpha_2$ with detuning $\delta_2=\omega_m$.  The converted output signal sideband emitted from cavity $\op{a}{}_2$ is beat against the transmitted control beam $\alpha_2$ on a high-speed photodiode.  The amplitude of the input signal tone near resonance with cavity mode $\op{a}{}_1$ is calibrated using a second reference signal generated through acousto-optic modulation of $\alpha_1$, whereas the amplitude of the output signal sideband is inferred from calibration of the control beam intracavity number $n_{c,2}$ and the optical transmission and detection chain (see Fig.~\ref{fig:EIT}a and Appendix).  Figure~\ref{fig:results}b plots the resulting power conversion efficiency, given by the magnitude squared of Eq.~(\ref{eqn:psd_expr}),  versus the ratio of $C_2/C_1$ as the power of control beam $\alpha_2$ is varied and for an input signal exactly resonant with cavity mode $\op{a}{}_1$ ($\Delta_1=\omega_m$).  The power conversion is referred directly to the input and output of the optomechanical cavity (i.e., not including additional losses in the optical link).  The solid red line shows the theoretical model for the conversion efficiency using the independently measured system parameters, and taking into account power-dependent effects on the optical losses in the silicon cavity modes (see Appendix).  Good correspondence is seen between the measured data and theory, with a maximum of the conversion occurring at $C_1 \approx C_2$ as expected from the matching condition.


The bandwidth of the wavelength conversion process is also probed, using matched conditions ($C_{1}=C_{2}\approx 16$), and as above, with control beam detunings $\delta_k=\omega_m$.  The modulation frequency generating $a_{in}$ is now varied from $(\Delta_1-\omega_m)/2\pi=-10~\mathrm{MHz}$ to $+10~\mathrm{MHz}$, sweeping the narrow-band input sideband signal across the transparency window of cavity mode $\op{a}{}_1$.  The corresponding measured output signal frequency shift follows that of the input signal ($\Delta_2=\Delta_1$), with the resulting power conversion efficiency plotted in the inset of Fig.~\ref{fig:results}b.  A peak efficiency, measured from the input to the output port at the optomechanical cavity system, is $\eta=2.2\%$, corresponding to an internal conversion efficiency of $\eta_{\text{max,int}}=93\%$.  Fitting the results to a Lorentzian as in Eq.~(\ref{eqn:skj}), we find the conversion bandwidth to be $1.55~\mathrm{MHz}$, equal to the optically damped mechanical linewidth of the mechanical resonance, $\gamma/2\pi$.  


\begin{figure}[t]
\begin{center}
\includegraphics[width=\columnwidth]{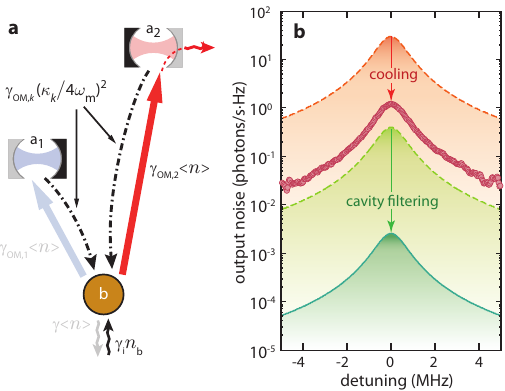}
\caption{\textbf{Output noise.} \textbf{a}, Schematic showing the relevant input noise terms contributing to the total noise at the output.  The rates of ``noisy'' phonon generation are $\gamma_i n_b$ due to the thermal bath and $\gamma_{\text{OM},(1,2)}(\kappa_{(1,2)}/4\omega_m)^2$ due to spontaneous Stokes scattering from the two control beams.  Cooling of the mechanical mode is also performed by the control beams, resulting in a theoretically cooled phonon occupancy of $\langle n \rangle=(1/\gamma)(\gamma_in_b+\sum_{j}\gamma_{\text{OM},j}(\kappa_{j}/4\omega_m)^2)$.  The output noise power is proportional to the output cavity cooling rate, $\gamma_{\text{OM},2}\langle n \rangle$. \textbf{b}, Output noise power spectral density.  The measured noise data (red circles) is shown along with several theoretical noise curves.  The dashed red curve is the corresponding output noise power in the absence of control beam cooling.  The solid (dashed) green curve is the output noise power due to spontaneous Stokes scattering in the presence (absence) of cavity filtering.}
\label{fig:Noise}
\end{center}
\end{figure}

The spectrum of the converted output signal for a series of signal frequencies $\Delta_1$ are shown in Fig.~\ref{fig:results}c.  As can be seen from this plot and the zoomed-in inset, the narrow-band converted photons sit atop a noise pedestal of bandwidth corresponding to the damped mechanical resonator.  This noise is quantified by the added noise quanta number component of Eq.~(\ref{eqn:psd_expr}).  As shown in the Appendix, under matched conditions the theoretical added noise is, 
\be
n_\text{added} = 2\eta^{-1}_1\left(\frac{\gamma_i n_b}{\gamma} + \frac{1}{2}\left(\frac{\kappa_1}{4\omega_m}\right)^2+\frac{1}{2}\left(\frac{\kappa_2}{4\omega_m}\right)^2\right),\label{eqn:nadded}
\ee
where the first term arises due to thermal noise of the cooled and damped mechanical resonator, and the last two terms are quantum noise resulting from the spontaneous scattering of the control beams (quantum back-action noise~\cite{Marquardt2007}).  Figure~\ref{fig:Noise}a pictorially indicates the various cooling, heating, and spontaneous scattering mechanisms that lead to the output noise of $\op{a}{}_2$.  A plot showing the measured output noise spectral density, calibrated in units of photon number, is given in Figure~\ref{fig:Noise}b for the optimal conversion efficiency of Fig.~\ref{fig:results}b.  From the peak of the output noise spectral density ($=\eta_1\eta_2 n_{\text{added}}$) in this plot, the added noise referred to the input of $\op{a}{}_1$ is estimated to be $n_\text{added}\approx60$ quanta.  The corresponding internal added noise (for $\eta=1$) is only $n_\text{added} \approx 6$, predominantly due to the cooled phonon occupation of the mechanical resonator ($\langle n \rangle \approx 3$).  Due to the strong filtering provided by the high-$Q$ cavities ($\kappa_k/\omega_m \ll 1$), the quantum back-action noise contributes an insignificant $4\times10^{-2}$ ($4\times10^{-3}$) added input (internal) noise photons.  The effectiveness of noise reduction in the optomechanical cavity system is highlighted in Fig.~\ref{fig:Noise}b by showing the estimated thermal output noise in the absence of sideband cooling by the control beams (dashed red curve) and the quantum back-action noise in absence of cavity filtering (green dashed curve).  


Although the demonstrated wavelength conversion in this work represents an important proof of principle, there are still considerable practical challenges to utilizing such cavity optomechanical devices for classical or quantum wavelength translation~\cite{Tanzilli2005,Kielpinski2011}.  With a thermal phonon occupancy $\langle n \rangle \leqslant \eta_1/2$ required for single photon conversion at unity signal-to-noise, a primary concern for quantum networking~\cite{Kimble2008} is the thermal energy stored in the mechanical resonator.  Although resolved sideband cooling of nanomechanical resonators to occupancies below one has recently been demonstrated~\cite{Teufel2011,Chan2011}, this relation emphasizes the additional need for efficient cavity coupling.  For chip-scale photonic devices similar to the one presented here, the technology already exists for efficient optical connectivity, with optical fiber-to-chip coupling of $\sim 95\%$~\cite{Bakir2010} and waveguide-to-cavity coupling efficiency of $\eta_1\gtrsim 0.999$ having been realized~\cite{Notomi2008}.  Further reduction in the output noise may also be realized by simply reducing the temperature of operation; for the $4$~GHz phonon frequency of the device in this study, $n_b \lesssim 10^{-3}$ for a bath temperature of $T\sim 10$~mK realizable in a helium dilution refrigerator.  Integration of cavity optomechanical devices into milliKelvin experiments is particularly relevant for superconducting quantum circuits, which themselves have already been strongly couped to mechanical resonators~\cite{OConnell2010}.  Addition of an optical cavity coupled directly, or via an additional acoustic waveguide, to the same mechanical resonator would provide a microwave-to-optical quantum interface~\cite{Safavi-Naeini2011b,Regal2011} similar to the optical-to-optical interface shown here.  In the case of solid-state qubits with optical transitions far from the telecommunication bands, such as the nitrogen-vacancy (NV) electron spin in diamond~\cite{Childress2006}, the radiation pressure nonlinearity of cavity optomechanics could be substituted for intrinsic materials nonlinearities in proposed cavity-based schemes for single photon wavelength conversion and pulse shaping~\cite{McCutcheon2009}.  Due to the excellent mechanical properties of diamond and recent advancements in thin-film diamond photonics~\cite{Faraon2011}, a cavity-optomechanical wavelength converter could be fabricated around an NV qubit from the same diamond host.  



\emph{Note:}  During the final preparation of this manuscript a similar worked appeared on the arXiv~\cite{Dong2012}.

\begin{acknowledgments}
This work was supported by the DARPA/MTO MESO program, Institute for Quantum Information and Matter, an NSF Physics Frontiers Center with support of the Gordon and Betty Moore Foundation, and the Kavli Nanoscience Institute at Caltech. JC and ASN gratefully acknowledge support from NSERC.
\end{acknowledgments}

\def\urlprefix{}
\def\url#1{}


\renewcommand{\thefigure}{A\arabic{figure}}
\renewcommand{\theequation}{A\arabic{equation}}
\setcounter{figure}{0}
\setcounter{equation}{0}

\appendix

\section{Theory of two-mode optomechanical system}


We begin by considering the Hamiltonian describing the optomechanical interaction between two distinct optical modes (indexed $k=1,2$) coupled to a shared mechanical mode with annihilation operator $\op{b}{}$,
\bea
\hat H &=& \sum_k \hbar\delta_k \opdagger{a}{k} \op{a}{k} +
\hbar\omega_m \opdagger{b}{} \op{b}{}+ (\op{b}{} + \opdagger{b}{}) \sum_k \hbar g_k \opdagger{a}{k} \op{a}{k}.\nonumber\\
\label{eqn:hamiltonian}
\eea
Each optical mode has a frequency $\omega_{k}$, and is driven by a laser at frequency $\omega_{l,k}$. The Hamiltonian above is written in the interaction picture with $\delta_{k} = \omega_{k}  - \omega_{l,k}$. 

As shown below, strong driving at a mechanical frequency red detuned from each cavity mode, $\delta_k \cong \omega_m$, causes an effective beam splitter interaction to take place between the mechanical mode and each optical cavity mode at an enhanced coupling rate $G_k = g_k |\alpha_k|$, where $\alpha_k$ is the square root of the photon occupation in optical mode $k$. As long as this coupling is weak with respect to the optical linewidths $\kappa_k$ ($G_k \ll \kappa_k$), an adiabatic elimination of the optical cavities results in new effective mechanical loss rates $\gamma_{\mathrm{OM},k}$ into each of the $k$ optical degrees of freedom in the system. This ``loss''  can provide an effective coupling between the optical cavity modes by allowing the exchange of excitations between the optical resonances through the mechanical motion of the system. We calculate exactly what these conversion rates are by using a scattering matrix formulation to understand the behaviour of the system. Though completely general expressions can be derived, to best understand the processes involved we focus on the parameter regime relevant to this experiment, i.e. the weak-coupling, sideband-resolved case where $\omega_m \gg \kappa \gg \gamma_\mathrm{OM}$, and follow a scattering matrix derivation of the induced optomechanical coupling between the optical waveguides coupled to each cavity mode.

The linearized Heisenberg-Langevin equations of motion are found using equation (\ref{eqn:hamiltonian}), the displacement $\op{a}{k} \rightarrow  \alpha_k$ + \op{a}{k}, and the inclusion of the optical input signal ($\op{a}{\mathrm{in},k}(t)$), optical vacuum noise ($\op{a}{\mathrm{i},k}(t)$), and mechanical thermal fluctuation fields ($\op{b}{\mathrm{in}}(t)$):
\bea
\dot{\op{b}{}}(t) &=& -\left(i\omega_{m} + \frac{\gamma_i}{2}\right) \op{b}{} - i\sum_k G_k (\op{a}{} + \opdagger{a}{})- \sqrt{\gamma_i}\op{b}{\mathrm{in}}(t)\nonumber\\
\dot{\op{a}{}}_k(t) &=& -\left(i\delta_k + \frac{\kappa_k}{2}\right) \op{a}{k} - iG_k(\op{b}{}+\opdagger{b}{}) \nonumber\\
&&-\sqrt{\kappa_{e,k}/2} \op{a}{\mathrm{in},k}(t) - \sqrt{\kappa^\prime_k}\op{a}{\mathrm{i},k}(t).\nonumber
\eea

The mechanical loss rate $\gamma_i=\omega_m/Q_m$, determines the coupling of the system to the thermal bath, and provides the most significant contribution in terms of noise processes relevant to the performance of the optomechanical wavelength converter. The other loss rates, $\kappa_k$, $\kappa_{e,k}$,  and $\kappa^\prime_k$ are, respectively for optical cavity mode $k$, the total optical loss rate, the optical loss rate associated with coupling into the waveguide $k$, and parasitic optical loss rate into all other channels that are undetected representing a loss of information. Due to the local evanescent side-coupling scheme studied here (which bi-directionally couples to the cavity mode), $\kappa_k = \kappa_{e,k}/2 + \kappa^\prime_k$, i.e. only a fraction $\eta_k  \equiv \kappa_{e,k}/2\kappa_k$ of the photons leaving the cavity get detected and $\eta_k \le 1/2$.

The equations of motion are linear and thus the system can be analyzed more simple in the frequency domain. Solving for the spectrum of the mechanical mode $\op{b}{}(\omega)$ in terms of the input noise operators, we find:
\bea
&\op{b}{}(\omega) =  \frac{-\sqrt{\gamma_i}\op{b}{\mathrm{in}}(\omega)}{i(\omega_m-\omega) + \gamma/2}~~~~~~~~~~~~~~~~~~~~~~~~~~~~~~~\nonumber\\
&+ \sum_k \frac{iG_k}{i(\delta_k -\omega) + \kappa_k/2} \frac{\sqrt{\kappa_{e,k}/2} \op{a}{\mathrm{in},k}(\omega) + \sqrt{\kappa^\prime_k} \op{a}{\mathrm{i},k}(\omega)}{i(\omega_m-\omega) + \gamma/2}\nonumber\\
&+ \sum_k\frac{iG_k}{-i(\delta_k + \omega) + \kappa_k/2} \frac{\sqrt{\kappa_{e,k}/2} \opdagger{a}{\mathrm{in},k}(\omega) + \sqrt{\kappa^\prime_k} \opdagger{a}{\mathrm{i},k}(\omega)}{i(\omega_m-\omega) + \gamma/2},\nonumber
\eea
where the mechanical frequency $\omega_m$ is now modified by the optical spring, and the mechanical linewidth is given by
$\gamma \equiv \gamma_i +\gamma_{\text{OM},1}+\gamma_{\text{OM},2}$. The optomechanical damping terms $\gamma_{\text{OM},k}$ come from coupling of the mechanical system to the optical mode $k$, and are given by the relation
\bea
\gamma_{\mathrm{OM},k} &&= 2|G_k|^2 \mathrm{Re} \Bigg[ \frac{1}{i(\delta_k-\omega_m)+\kappa_k/2} \nonumber\\
&&- \frac{1}{-i(\delta_k+\omega_m)+\kappa_k/2} \Bigg]\nonumber\\
&&= \frac{4 |G_k|^2}{\kappa_k}~~~\text{for}~\delta_k\approx\omega_m.\label{eqn:damping}
\eea
The last expression is a simplification that is often made, and is equivalent to looking at spectral properties at detunings from the mechanical frequency much smaller than $\kappa$ (so the optical lineshape isn't taken into account). Under this approximation, the optical spectrum is
\bea
\frac{\kappa_k}{2}\op{a}{k}(\omega) &=  \frac{i G_k \sqrt{\gamma_i}\op{b}{\mathrm{in}}(\omega)}{i(\omega_m-\omega) + \gamma/2}~~~~~~~~~~~~~~~~~~~~~~~~~~~~~~~\nonumber\\
&+\sum_j \frac{2G_jG_k}{\kappa_j}\frac{\sqrt{\kappa_{e,j}/2} \op{a}{\mathrm{in},j}(\omega) + \sqrt{\kappa^\prime_j} \op{a}{\mathrm{i},j}(\omega)}{i(\omega_m-\omega) + \gamma/2}\nonumber\\
&+\sum_j \frac{iG_jG_k}{2\omega_m}\frac{\sqrt{\kappa_{e,j}/2} \opdagger{a}{\mathrm{in},j}(\omega) + \sqrt{\kappa^\prime_j} \opdagger{a}{\mathrm{i},j}(\omega)}{i(\omega_m-\omega) + \gamma/2}\nonumber \\
& - \sqrt{\kappa_{e,k}/2} \op{a}{\mathrm{in},k}(\omega) - \sqrt{\kappa^\prime_k} \op{a}{\mathrm{i},k}(\omega).
\label{eqn:a_w}
\eea
From this expression, we see that there are several noise operators incident on optical mode $k$. The thermal fluctuations from the environment, $\op{b}{\mathrm{in}}(\omega)$, are converted into noise photons over the mechanical bandwidth $\gamma$. There is also an induced coupling to the optical mode $j\neq k$, and photons originally incident only on mode $j$, i.e. $\op{a}{\mathrm{in},j}(\omega)$ are now also coupled into $\op{a}{k}$. Finally we note that vacuum noise creation operators are also present in this expression. These give rise to quantum noise through spontaneous emission of phonons, though we show below that as $\omega_m$ is made very large with respect to $\kappa_k$, these terms diminish in importance.

From outside the optomechanical system, we only have access to photons sent into the system, $\op{a}{\mathrm{in},j}$, and those leaving the system, $\op{a}{\mathrm{out},k}$, on transmission. The relation between these operators is best understood through scattering parameters, and can be derived using the equation for $\op{a}{k}(\omega)$ (Eq.~\ref{eqn:a_w}) and the input-output boundary conditions $\op{a}{\mathrm{out},k}(\omega) =\op{a}{\mathrm{in},k}(\omega) +\sqrt{\kappa_{e,k}/2} \op{a}{k}(\omega)$.  After some algebra, the output operator is expressed as
\bea
\op{a}{\mathrm{out},k}(\omega) &=& s_{\mathrm{th},k}(\omega) \op{b}{\mathrm{in}}(\omega) \nonumber\\&&
+ t_k(\omega)\op{a}{\mathrm{in},k}(\omega) + s_{kj}(\omega) \op{a}{\mathrm{in},j}(\omega)\nonumber\\&&
+\sum_m n_{\text{opt},m}(\omega) \op{a}{\mathrm{i},m}(\omega)\nonumber
\\&&
+\sum_m s_{\text{adj,in},m}(\omega) \opdagger{a}{\mathrm{in},m}(\omega)\nonumber
\\&&
+\sum_m s_{\text{adj,i},m}(\omega) \opdagger{a}{\mathrm{i},m}(\omega)
\label{eqn:full_scattering_relation}
\eea
The scattering coefficient $s_{\mathrm{th},k}(\omega)$ is the conversion efficiency of mechanical thermal noise to photons, and is given by
\bea
s_{\mathrm{th},k}(\omega) = i \sqrt{\eta_k} \frac{\sqrt{\gamma_i \gamma_{\mathrm{OM},k}}}{i(\omega_m - \omega)+\gamma/2}.
\eea
The coefficient $t_k(\omega)$ in our system is transmission amplitude, given by
\bea
t_k(\omega) = \left( 1 - 2\eta_k \right) + 
\eta_k\frac{\gamma_{\text{OM},k}}{i(\omega_m - \omega) + \gamma/2}.
\eea
In principle, this coefficient is the electromagnetically induced transparency transmission coefficient, though here it is written about the mechanical frequency for detunings on the order of $\gamma$ and the $\kappa$'s are too large ($\kappa \gg \gamma$) for the optical lineshape to be considered in the expression. This is the expected result of the weak coupling approximation. Finally, the most important coefficient is the wavelength conversion coefficient $s_{kj}(\omega)$ which is given by
\bea
s_{kj}(\omega) = \sqrt{\eta_k \eta_j}  \frac{\sqrt{\gamma_{\text{OM},k}\gamma_{\text{OM},j}}}{i(\omega_m - \omega) + \gamma/2}.
\label{eqn:skj_app}
\eea

\subsection{Efficiency,  Bandwidth, and Noise}

We calculate the spectral density of the output field on port $k$ ($S_{\mathrm{out},k}(\omega)$) assuming an input field spectral density on the opposing optical port $j$ ($S_{\mathrm{in},j}(\omega)$), vacuum inputs on all other optical channels, and thermal noise from a phonon bath with thermal occupation $n_b$. The spectral densities here have units of $\text{photons}/\text{Hz}\cdot\text{s}$ and can be interpreted as photon flux per unit bandwidth. At first we ignore the field creation operators in the scattering relation (\ref{eqn:full_scattering_relation}) (which give rise to quantum noise and can be neglected when the mechanical resonator occupation number is greater than one) and arrive at the following expression:
\bea
S_{\mathrm{out},k}(\omega) &=& \int_{-\infty}^{\infty} \mathrm{d}\omega^\prime~ \avg{\opdagger{a}{\mathrm{out},k}(\omega)\op{a}{\mathrm{out},k}(\omega^\prime)} \nonumber\\
 &=& \int_{-\infty}^{\infty} ~ 
s^\ast_{\mathrm{th},k}(-\omega) s_{\mathrm{th},k}(\omega^\prime) \avg{\opdagger{b}{\mathrm{in}}(\omega)\op{b}{\mathrm{in}}(\omega^\prime)}\nonumber\\&&+
t^\ast_k(-\omega) t_k(\omega^\prime) \avg{\opdagger{a}{\mathrm{in},k}(\omega)\op{a}{\mathrm{in},k}(\omega^\prime)}\nonumber\\&&+
s^\ast_{kj}(-\omega) s_{kj}(\omega^\prime) \avg{\opdagger{a}{\mathrm{in},j}(\omega)\op{a}{\mathrm{in},j}(\omega^\prime)}\nonumber\\&&+
n^\ast_{\text{opt},k}(-\omega) n_{\text{opt},k}(\omega^\prime) \avg{\opdagger{a}{\mathrm{i},k}(\omega)\op{a}{\mathrm{i},k}(\omega^\prime)}\nonumber\\&&+
n^\ast_{\text{opt},j}(-\omega) n_{\text{opt},j}(\omega^\prime) \avg{\opdagger{a}{\mathrm{i},j}(\omega)\op{a}{\mathrm{i},j}(\omega^\prime)} \mathrm{d}\omega^\prime\nonumber.
\eea
This reduces to,
\bea
S_{\mathrm{out},k}(\omega) &=& \eta_k \frac{\gamma_i \gamma_{\text{OM},k}}{(\omega + \omega_m)^2 + (\gamma/2)}  n_b \nonumber\\&&
+ \eta_k \eta_j \frac{\gamma_{\text{OM},k} \gamma_{\text{OM},j}}{(\omega + \omega_m)^2 + (\gamma/2)^2} S_{\mathrm{in},j}(\omega),\nonumber
\eea
accounting for the noise autocorrelation functions. From here, we see that the maximum photon wavelength conversion efficiency is given by 
\be
\eta_\text{max} = \eta_1 \eta_2 \frac{4 \gamma_{\text{OM},1} \gamma_{\text{OM},2}}{(\gamma_i + \gamma_{\text{OM},1} + \gamma_{\text{OM},2})^2}.
\ee
The conversion process has a bandwidth (BW) equal to the mechanical linewidth
\be
\text{BW} \equiv \gamma_i + \gamma_{\text{OM},1} + \gamma_{\text{OM},2}.
\ee
From the same expression, we calculate a value for the optical signal to noise ratio (OSNR), to find
\be
\text{OSNR}^\text{classical}_{kj} = \eta_j \frac{\gamma_{\text{OM},j}}{\gamma_i n_b} S_{\mathrm{in},j}(\omega). \label{eqn:OSNR_classical}
\ee

Equation~(\ref{eqn:OSNR_classical}) only takes into account the thermal noise in the system and therefore holds only when $\gamma_i n_b / \gamma_{\text{OM},j} \gg 1$. At higher powers, the quantum noise processes which give rise to the back-action limit in cooling also give rise to excess noise for wavelength conversion. To consider these terms, the creation operators in Eq. (\ref{eqn:full_scattering_relation}) must not be neglected and the full relation is then found to be
\bea
S_{\mathrm{out},k}(\omega) &=& \eta_k \frac{ \gamma_{\text{OM},k}}{(\omega + \omega_m)^2 + (\gamma/2)^2}  \gamma_i n_b \nonumber\\&&
  + \eta_k \frac{ \gamma_{\text{OM},k}}{(\omega + \omega_m)^2 + (\gamma/2)^2} \gamma_{\text{OM},k} \left(\frac{\kappa_k}{4\omega_m}\right)^2 \nonumber\\&&
 + \eta_k \frac{ \gamma_{\text{OM},k}}{(\omega + \omega_m)^2 + (\gamma/2)^2}  \gamma_{\text{OM},j} \left(\frac{\kappa_j}{4\omega_m}\right)^2  \nonumber\\&&
+ \eta_k \eta_j \frac{\gamma_{\text{OM},k} \gamma_{\text{OM},j}}{(\omega + \omega_m)^2 + (\gamma/2)^2} S_{\mathrm{in},j}(\omega).
\eea
The quantum limited $\text{OSNR}$ is then found to be
\bea
\text{OSNR}_{kj} &=  \frac{\eta_j\gamma_{\text{OM},j}{S_{\mathrm{in},j}(\omega)}}{\gamma_i n_b + \gamma_{\text{OM},k} \left(\frac{\kappa_k}{4\omega_m}\right)^2 + \gamma_{\text{OM},j} \left(\frac{\kappa_j}{4\omega_m}\right)^2}.
\eea
For the quantum back-action terms to become significant as compared to the thermal noise on the mechanical system, one needs $\gamma_i n_b / \gamma \sim  \left(\frac{\kappa}{4\omega_m}\right)^2$. This is a regime that is yet to be reached in experiments and as such the quantum back-action noise terms can be neglected for experiments to date. Perhaps more importantly, we ask ``what is the amount of noise added to a signal of a single photon?''. We can express the signal to noise ratios as $\text{OSNR}_{kj} =  S_{\mathrm{in},j}(\omega) / n_\text{added}$, with
\bea
n_\text{added} &=& \frac{\gamma_i n_b + \gamma_{\text{OM},k} \left(\frac{\kappa_k}{4\omega_m}\right)^2 + \gamma_{\text{OM},j} \left(\frac{\kappa_j}{4\omega_m}\right)^2}{\eta_j \gamma_{\text{OM},j}}.\nonumber \\
\eea
Assuming $\kappa_j = \kappa_k = \kappa$, $\gamma_i \ll \gamma$, and $\gamma_{\text{OM},j} =\gamma_{\text{OM},k} = \gamma_{\mathrm{OM}}$, this expression becomes 
\be
n_\text{added} \approx 2\eta^{-1}_j\left(\frac{\gamma_i n_b}{\gamma} + \left(\frac{\kappa}{4\omega_m}\right)^2\right).
\ee
To achieve the threshold where a single photon incoming on port $j$ is converted, and is more likely to be detected on port $k$ than an upconverted thermal excitation, we require $n_\text{added} < 1$. This requirement is equivalent to stating that in the absence of signals on port $j$ and $k$, a cooled phonon occupation of $\bar{n} = \eta_j/2$ must be achieved.  We note that $\eta_k$ does not make an appearance in this equation, since reduced output coupling efficiency attenuates the thermal noise and signal equivalently. For the same reason, $1/\eta_j$ is present in the noise quanta expression. The factor of two can be understood to come from the fact that for good conversion, we require the photon to enter, and exit the system before a phonon is up converted. The former two processes happen at $\gamma_\text{OM}$, while the latter is given by the thermalization rate $\gamma_i n_b$.

\begin{figure*}[t!]
\begin{center}
\includegraphics[width=1.8\columnwidth]{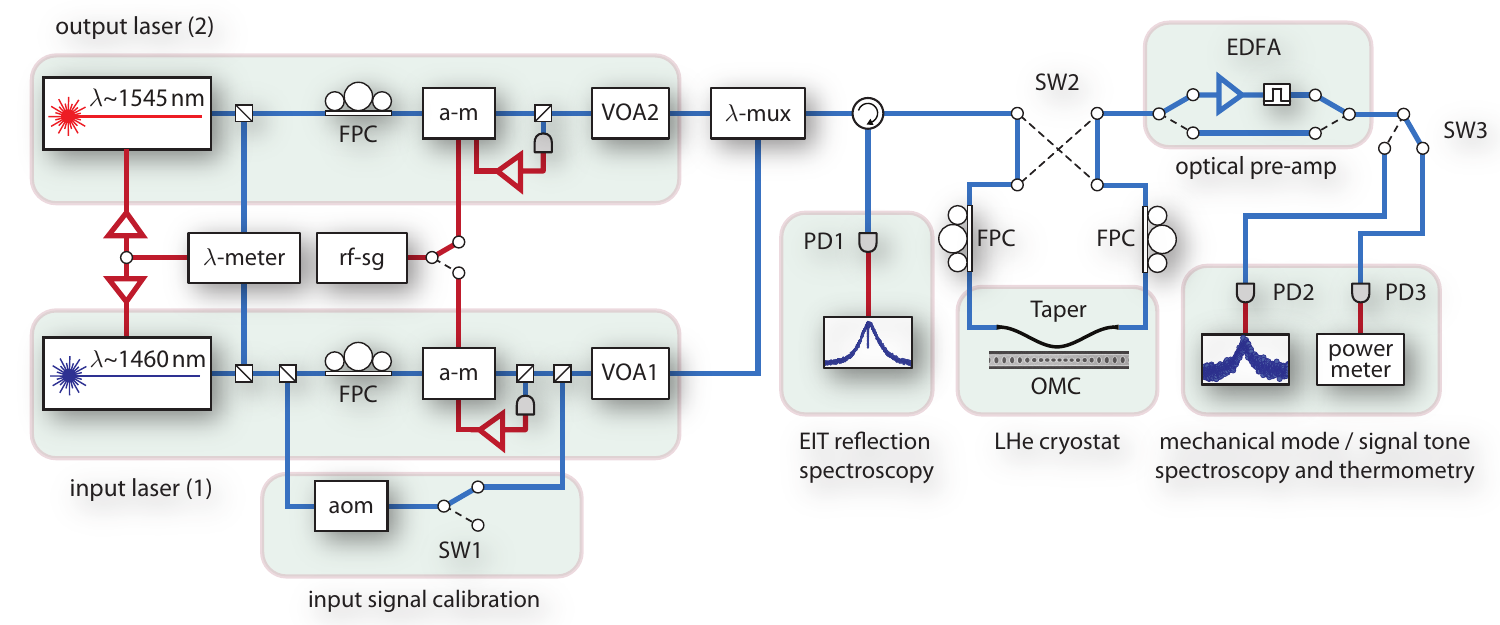}
\caption{\textbf{Detailed schematic of the experimental setup.} Two tunable, external cavity diode lasers (ECDL) are used as control laser beams used to drive the wavelength conversion process.  The wavelength of both lasers is monitored and locked to an absolute frequency of better than $\pm 5$~MHz using a wavemeter ($\lambda$-meter).  Both control beams can be amplitude modulated (a-m) using an RF signal generator (rf-sg), which produces the necessary optical sidebands to perform EIT-like spectroscopy (see main text) and to generate the input signal for the wavelength conversion process.  An acousto-optic modulator (aom) is used to calibrate the input signal (switching in the calibration signal with SW1).  The light from both lasers are attenuated (VOA1, VOA2) to the desired power, combined using a wavelength multiplexer ($\lambda$-mux), and then sent into the dimpled optical fibre taper which is used to optically couple to the optomechanical crystal.  The optomechanical crystal sample and fibre taper are both located in a continuous-flow liquid helium cryostat.  Before entering the cryostat, the light passes through fibre polarization controllers (FPC) to align the polarization of the light with that of the predominantly in-plane linear polarization of the optomechanical crystal cavity modes.  The light also passes through an optical switch (SW2), allowing the light to be switched between either direction of propagation through the fibre taper, and providing an accurate calibration of the losses in the taper before and after the optomechanical crystal cavity (measuring the input-power-dependent optomechanical damping in both transmission directions allows one to separate the total loss of the taper into losses before and after the cavity).  The transmitted light from the cavity is either optically pre-amplified using an erbium doped fibre amplifier (EDFA) and sent to a high-speed photodetector (PD2) for measurement of the microwave power spectrum, or using optical switch (SW3), sent to an optical power meter (PD3, power meter) for power calibration.}
\label{fig:si_setup}
\end{center}
\end{figure*}

\section{Calibration of Wavelength Conversion Efficiency}
\subsection{Generation and conversion of the input signal}
We begin by considering the input signal to the first optical cavity mode.
This signal is generated by weakly modulating the $\alpha_1$ control beam (represented by amplitude $A_1$) at a frequency $\Delta_1$ using an electro-optic modulator (EOM) producing
\bea
A_1 & \to& A_1 + \frac{\beta_1}{4} A_1 e^{-i \Delta_1 t} + \frac{\beta_1}{4} A_1 e^{i \Delta_1 t}e^{i\phi},
\label{eqn:A1}
\eea
where $\beta_1$ is the modulation index of the input signal and $\phi$ accounts for any phase difference between the sidebands ($\phi=0$ if there is pure amplitude modulation).
The input sideband at frequency $\omega_{l,1} + \Delta_1$ is nearly resonant with the first cavity mode at $\omega_1$, whereas the lower frequency sideband at $\omega_{l,1} - \Delta_1$ is detuned by approximately $-2\omega_m$.  As such, only the upper frequency sideband is resonant with the transparency window of the first cavity mode, and only this sideband is converted by the optomechanical crystal cavity into a sideband at the second, output cavity mode $\op{a}{}_2$. The converted sideband is generated from the control laser beam $\alpha_2$ of the second cavity mode with frequency $\omega_{l,2} + \Delta_1$.  Thus, the output field amplitude near the frequency of the second control laser beam (amplitude $A_2$) is, ignoring $\phi$,
\be
A_2 \to A_2 + s_{21}(\Delta_1) \frac{\beta_1}{4} e^{i \Delta_1 t}.
\ee
The power measured by a photodetector from the optical signal eminating from the output cavity mode is proportional to  
\be
P_2 = |A_2|^2 \left( 1 + |s_{21}(\Delta_1)| \frac{\beta_1 A_1}{4 A_2} \cos \Delta_1 t  \right).
\ee
We thus define $\beta_2 =  |s_{21}(\Delta_1)| \beta_1 A_1/4 A_2$ as the modulation index of the output signal, 
\be
P_2 = |A_2|^2 \left( 1 + \beta_2 \cos \Delta_1 t \right).
\ee
By careful calibration of $\beta_1$ and $\beta_2$, and measurement of the control beam power ($P_{\alpha_k}$), the conversion scattering matrix element can be determined:
\begin{equation}
s_{21}(\Delta_1) = 2 \frac{\beta_2 A_2}{\beta_1 A_1} = 2 \frac{\beta_2}{\beta_1}\sqrt{\frac{P_{\alpha_2}}{P_{\alpha_1}}}.
\end{equation}

\subsection{Calibration of the Input Signal}

\begin{figure}[ht!]
\begin{center}
\includegraphics[width=\columnwidth]{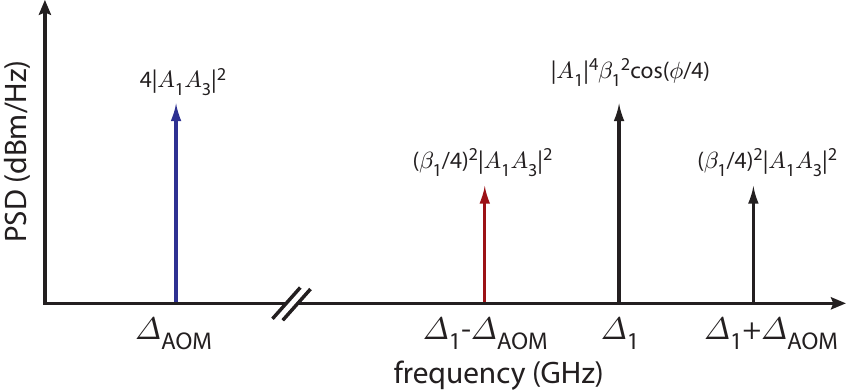}
\caption{\textbf{Calibration tones.} Schematic of the four tones appearing in the electronic power spectrum of the photocurrent generated by the optical input signal and the additional calibration signal.  The three additional tones proportional to $A_3$ are generated to calibrate the input signal modulation index, $\beta_1$.}
\label{fig:si_1450_calib}
\end{center}
\end{figure}

The appearance of $\phi$ in equation (\ref{eqn:A1}), along with its sensitivity to effects such as chromatic dispersion in the optical fiber and components, make calibration of $\beta_1$ more challenging due to interference between the upper and lower frequency sidebands. Direct photodetection of the optical input signal results in a $\phi$-dependent measured beat signal between the carrier and the two sidebands. To bypass this problem, we generate an additional single sideband using an acousto-optic modulator (AOM), and beat it against each of the input optical sidebands. This is accomplished by splitting off a portion of the optical input signal prior to creating the EOM sidebands, and frequency shifting it by $\Delta_\text{AOM}$ (see Fig.~\ref{fig:si_setup}). This AOM-shifted signal is then recombined with the main input signal, giving an overall signal equal to
\be
A_{\text{out},1} = A_1 + \frac{\beta_1}{4} A_1 e^{-i \Delta_1 t} + \frac{\beta_1}{4} A_1 e^{i ( \Delta_1 t + \phi )} + A_3 e^{i \Delta_\text{AOM} t},
\ee
where $A_3$ is the field amplitude of the signal split off from the input signal prior to the EOM. The total photodetected signal is then given by
\bea
|A_{\text{out},1}|^2&=& |A_1|^2 + |A_3|^2 \nonumber \\&&
+ |A_1|^2 \beta_1 \cos(\Delta_1 t + \phi/2) \cos(\phi/2)\nonumber \\&&
 + 2 |A_1 A_3| \cos(\Delta_\text{AOM} t)  \nonumber \\&&
 +|A_1 A_3| \frac{\beta_1}{2} \cos[(\Delta_1-\Delta_\text{AOM})t]\nonumber \\&&
 + |A_1 A_3| \frac{\beta_1}{2} \cos[(\Delta_1+\Delta_\text{AOM})t + \phi] .
\eea
From this equation we see that the optical power at the detector consists of components at zero frequency (DC) as well as four modulated tones (see Fig.~\ref{fig:si_1450_calib}). By taking the ratio of the tone at $\Delta_{\text{AOM}}$ and $\Delta_1 - \Delta_{\text{AOM}}$, the value of $\beta_1$ can be accurately determined, independent of $\phi$.  Note, that in order to determine $\beta_2$ no such additional calibration signal is required.  The filtering properties of the input and output cavities results in a single output sideband, with no phase dependence of the photodetected intensity spectrum.  In this case, careful optical and electronic calibration of the measured output cavity ($\op{a}{}_2$) optical transmission intensity, similar to that described in the Supplementary Information of Ref.~\cite{Chan2011},  provides an accurate estimate of $\beta_2$ and the wavelength converted signal strength.

\subsection{Power dependent optical cavity loss}

\begin{figure}[ht!]
\includegraphics[width=1\columnwidth]{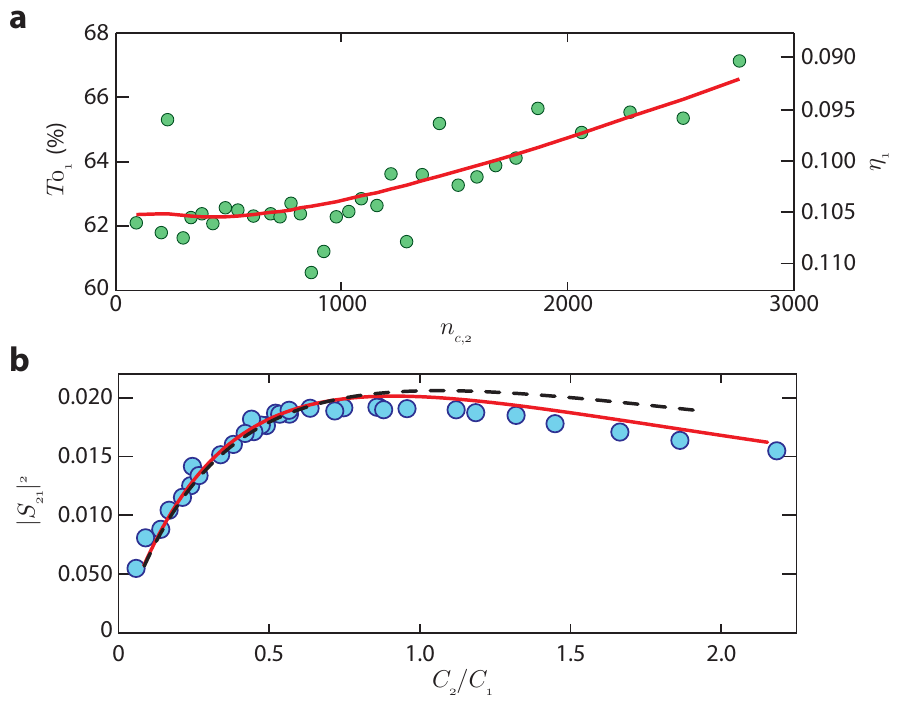}
\caption{\textbf{Effect of large intracavity photon population.} \textbf{a}, Recorded coupling depth of the first order optical cavity as a function of the intracavity photon population of the second order mode (green circles).  The black curve represents a fit to the extracted coupling depth. \textbf{b}, Conversion efficiency as a function of $C_2/C_1$.  The dashed black line represents the theoretical values from the nominal system parameters while the solid red line takes into account the effect of the large intracavity photon population of cavity mode 2 ($n_{c,2}$).  The blue circles are the extracted data points and show much better correspondence to theory when this effect is accounted for.}
\label{fig:si_T0}
\end{figure}

To model the wavelength conversion process, all of the optical cavity mode and mechanical resonator parameters entering into Eq.~\ref{eqn:skj_app} must be independently determined. Most of these parameters are measured as described in the main text; however, additional measurements were performed to determine the power dependent optical cavity loss in the silicon optomechanical crystal device.  Due to two-photon absorption in silicon, the parasitic optical cavity loss of both cavity modes is not static, but rather depends upon both control beam powers.  This effect is particularly acute for the higher $Q$-factor first-order cavity mode ($\op{a}{}_1$), where in Fig.~\ref{fig:si_T0}a we plot the normalized on-resonant optical transmission ($T_{0,1}$) of the first-order cavity mode versus the power (as represented by the intra-cavity photon population, $n_{c,2}$) of control beam $\alpha_2$ feeding the second-order cavity mode.  The steady rise in the on-resonance transmission versus $n_{c,2}$ is attributable to the increased optical absorption loss stemming from free-carriers generated by two-photon absorption of the control beam $\alpha_2$.  The corresponding change in coupling efficiency, $\eta_1 \equiv \kappa_{e,1}/2\kappa_1$, is shown on the right-side axis of Fig.~\ref{fig:si_T0}a.  Neglecting the effects of nonlinear optical absorption and free carriers on the intrinsic optical quality factor of $\op{a}{}_1$ leads to a theoretical model of conversion efficiency denoted by the dashed black line in Fig~\ref{fig:si_T0}b.  Taking into the account the nonlinear optical loss measured in Fig.~\ref{fig:si_T0}a leads to the theoretical conversion efficiency denoted by the solid red line, showing much better correspondence with the data (the solid red line is the model fit shown in the main text).  The effect of two-photon absorption and free-carrier absorption is much smaller on the second-order cavity mode ($\op{a}{}_2$) due to its already lower optical $Q$-factor, and we neglect it here.  The control-beam-generated free carriers also affect the intrinsic mechanical quality factor ($\gamma_i$) as was shown in Ref.~\cite{Chan2011}, however, this affect is small for the control beam powers studied here (the efficiency curve of Fig~\ref{fig:si_T0}b is also not modified as we plot it versus the ratio of cooperativities for which $\gamma_i$ cancels out). 

\end{document}